# GROOT: GENERAL-PURPOSE AUTOMATIC PARAMETER TUNING ACROSS LAYERS, DOMAINS, AND USE CASES

Robert Krahn  
*TU Dresden*

Josia Mädler  
*TU Dresden*

Christoph Seidl  
*TU Dresden*

Christof Fetzer  
*TU Dresden*

**ABSTRACT**

Modern software systems are executed on a runtime stack with layers (virtualization, storage, trusted execution, etc.) each incurring an execution and/or monetary cost, which may be mitigated by finding suitable parameter configurations. While specialized parameter tuners exist, they are tied to a particular domain or use case, fixed in type and number of optimization goals, or focused on a specific layer or technology. These limitations pose significant adoption hurdles for specialized and innovative ventures (SIVs) that address a variety of domains and use cases, operate under strict cost-performance constraints requiring tradeoffs, and rely on self-hosted servers with custom technology stacks while having little data or expertise to set up and operate specialized tuners.

In this paper, we present Groot – a general-purpose configuration tuner designed to a) be explicitly agnostic of a particular domain or use case, b) balance multiple potentially competing optimization goals, c) support different custom technology setups, and d) make minimal assumptions about parameter types, ranges, or suitable values. Our evaluation on both real-world use cases and benchmarks shows that Groot reliably improves performance and reduces resource consumption in scenarios representative for SIVs.

## 1. INTRODUCTION

A modern software system is executed on a runtime stack with layers, e.g., hardware, hypervisor, virtualization runtime (e.g., containers / VMs), runtime framework (e.g., JVM, Python) or Trusted Execution Environment ( e.g., Intel SGX, ARM TrustZone). On top of the runtime stack runs a binary we refer to as an application (one for single systems, multiple for distributed systems). Adopting new technology or continuous application development may lead to changes to the runtime stack, e.g., by adding new layers for TEEs.

Companies using large data center services and innovators alike seek to improve performance of their software system while keeping operation cost in check, e.g., by tuning application parameters or optimizing resource allocations. In principle, each layer may be configured through parameters that influence the application's performance. A misconfiguration, even on a layer further below, has the potential to limit the application's performance, e.g., by providing too few CPU cores or too little RAM on the virtualization layer.

Automatic tuning aims to find suitable configurations of parameter values for improving application performance and is challenging due to: *Parameter complexity* – The number and possible combinations of parameter values for all layers creates a complex search space for tuning; *Cross-Layer Effects* – Changing a parameter value on a lower level may influence performance on an upper layer; *Multiple Objectives* – The optimization procedure may be geared towards more than one goal, e.g., improving performance while reducing cost; *Tuning Boundaries* – The optimization procedure may be confined to certain fixed limits, e.g., respecting a maximum latency specified in a service-level agreement (SLA); *Parameter Changeability* – Parameter values may be changed either before application start (offline) or during application runtime (online).

While large corporations with long-running systems face few constraints due to abundant resources and stable workloads, current automatic tuners pose major adoption hurdles for specialized and innovative ventures (SIVs) such as SMEs, R&D labs, and start-ups. These organizations often rely on self-hosted servers instead of large-scale cloud deployments to maintain custom stacks, operate under strict cost-performance constraints, lack data or expertise, require rapid and frequent adoption of new software and runtime stacks to stay competitive, and build prototypes across diverse technologies where finding dedicated tuners deters from innovation.

In consequence, we argue that existing tuners with their focus on individual challenges, selected use cases or specific technologies are not sufficient for SIVs. At the same time, devising a custom-tailored tuner for each specific use case and combination of technologies is infeasible not only due to the incurred cost in terms of both money and time but also the lack of knowledge regarding tuning both on the specific scenario and at the respective companies. To make automatic parameter tuning available to SIVs, we propose the following requirements for a general-purpose tuner:

**R1 – Multiple Layers:** Must be able to process parameters on multiple layers of the runtime stack.
**R2 – Multiple Objectives:** Must support competing objectives with parameter thresholds.
**R3 – Offline/Online:** Must support both offline and online tuning.
**R4 – Adaptable:** Must be adaptable to different setups and changes of runtime layers.
**R5 – Agnostic:** Must be agnostic to the specific technology and use case of a software system.

Guided by these requirements, we developed Groot, a general-purpose parameter tuner facilitating automatic performance tuning for applications of SIVs with their wide variety of use cases and plethora of technology choices.

## 2. BACKGROUND

Optimizing a system's performance depends on the interplay between parameters, metrics, and objectives across its layers. Parameters exposed by a layer allow its configuration, e.g., to alter resource provisioning, which may affect performance of higher layers and the application (e.g., shared_buffers for Postgres or memory limits for a Docker container). A parameter configuration is a concrete assignment of values to a set of parameters. A metric quantifies observable system qualities (e.g., measuring latency) and may have labels with structured metadata used for filtering, normalization and prioritization during performance tuning.

A performance tuner seeks to improve a system's runtime qualities (e.g., throughput) while reducing execution cost in terms of resource consumption and/or monetary expenditures. A tuning objective defines a characteristic to tune and its desirable tuning direction (e.g., increase throughput). A threshold constrains tuning of a characteristic's value within a boundary (e.g., response time always below 10 ms). Multi objective tuning considers multiple, potentially competing, objectives (e.g., throughput and monetary cost) where a weight governs relative importance of an individual target.

A tuning algorithm determines candidate configurations promising a performance improvement by analyzing a system state (currently observed metrics and active parameter configuration) and systematically exploring a search space (the Cartesian product of relevant parameters, their value ranges and their associated metrics). A candidate configuration is enacted either online (without restarting a layer) or offline (restarting a layer and those above) for changes to take effect before the performance tuning loop starts the next iteration.

## 3. STATE OF THE ART

A host of related work exists on automated configuration tuning and performance optimization but a research gap for a general-purpose configuration tuner remains unaddressed as existing work addresses only subsets existing challenges. We discuss state of the art grouped by characteristics:

**Targeted Layer(s)**
*Single-layer tuners* apply changes only at a specific layer of the runtime stack without considering other layers. GPTuner (Lao et al., 2025), K2vTune (Lee et al., 2024), CDBTune (J. Zhang et al., 2019) and Hunter (Cai et al., 2022) operate on the DB layer. Ant (Cheng et al., 2016), DynamiConf (Gounaris et al., 2017), and XDataExplorer (Guo et al., 2022) tune only parameters specific to scheduling of MapReduce/Spark. SGXTuner (Mazzeo et al., 2022) tunes enclave runtime parameters of TEEs.
*Multi-layer tuners* enact parameter values on multiple layers of the runtime stack acknowledging cross-stack dependencies, e.g. resource allocation (OS-layer) and performance (application layer). PipeTune (Rocha et al.,

2020) simultaneously tunes OS related parameters and hyperparameters for DNN training. ResTune (X. Zhang et al., 2021) and CGPTUNER (Cereda et al., 2021) tune on the virtualization along with DB layer. OPPerTune (Somashekar et al., 2024) tunes cluster manager parameters and hardware resource allocations for container deployments.

Single-layer tuners neglect interactions between different layers; multi-layer tuners acknowledge these interactions but consider a fixed set or type of layers. Accordingly, existing tuners remain tailored to predefined settings, which ultimately prevents them from generalizing to more diverse stack compositions.

**Tuning Guidance**
*Single-objective tuning* tunes parameter configurations with a single objective, primarily focused on various aspects of performance, e.g., throughput (ConfAdvisor (Chen et al., 2021), SGXTuner), DNN error rates (AutoKeras (Jin et al., 2019)), job execution time (Gunther, Ant, LOCAT (Xin et al., 2022)), Latency (OtterTune (Van Aken et al., 2017)), with other tuners capable of switching between objectives but optimizing one at a time (SelfTune (Karthikeyan et al., 2023), OPPertune, ByteScheduler (Peng et al., 2019), Ituned (Duan et al., 2009)). CGPTuner extends single objective tuning with penalizing parameter configurations in violation of constraints on secondary qualities. Single-objective tuning neglects the complexity of real-world systems, where multiple objectives and cross-layer interactions must be addressed simultaneously.
*Multi-objective tuning* optimizes towards multiple potentially competing goals that have to be weighed or prioritized, e.g., PipeTune, Vizier (Golovin et al., 2017), and BestConfig (Zhu et al., 2017).

**Change Enactment**
*Offline tuning* enacts parameter changes and restarts the targeted layer, generally because changes take effect only after restarts, e.g., setting environment variables (SGXTuner) or creating static configurations (ConfAdvisor), or because tuning results are measurable only after termination (PipeTune, Gunther, Ant, or MALIBOO (Guindani et al., 2022)). Hybrid solutions execute offline tuning on trials to configure online services (Metis (Li et al., 2018)).
*Online tuning* enacts parameter changes without restarts if parameters and target layer permit to keep the system in operation. Examples are OPPerTuner, Hunter, SelfTune, and ByteScheduler (partially).
While each approach is effective within its scope, a generic tuner needs to accommodate for both offline and online parameter change enactment to be applicable and adaptable to various use cases.

**Specialization**
*Domains and use cases* allow specialized tuners to exploit peculiarities of the typical usage profiles, application parameters, or software architecture. These domains include databases (OtterTune, ResTune, CDBTune, Hunter), container deployment (ConfAdvisor), Deep Neural Networks (DNN) (PipeTune, ByteScheduler, AutoKeras (Jin et al., 2019)), Trusted Execution Environments (SGXTuner), and distributed processing (Gunther, Starfish (Herodotou et al., 2011)).
*Model-based tuning* algorithms construct predictive models of the system to estimate performance outcomes and guide parameter selection, reducing the need for exhaustive trials, e.g., OPPerTune, OtterTune, Flash (Nair et al., 2018), UDAO (Zaouk et al., 2019), Maliboo, and LOCAT. *Search-based tuning* algorithms explore configurations through direct experimentation to empirically observe performance, without relying on prior models, e.g., Ant, BestConfig, Starfish, SGXTuner. Specialization by domain or use case can improve performance in a narrow setting but restricts broader applicability. Model-based tuning requires a preliminary modeling step, yet the resulting abstraction may not adequately capture the irregular and rugged nature of system behavior during tuning. Search-based methods, by acting solely on observed performance without assumptions, can achieve robust tuning results provided they are based on an appropriate algorithm.

**Representative Prior Work**
BestConfig and Vizier are examples for use case and domain-agnostic approaches to automatic configuration tuning. BestConfig focuses on multi-objective constrained optimization. Vizier focuses on generality by providing a library service entailing tuning algorithms. While both BestConfig and Vizier contribute towards general applicability of tuning, they fall short in critical aspects: BestConfig lacks cross-layer tuning and Vizier requires high integration efforts. Groot is designed to tune for multiple objectives across multiple layers while providing all necessary components for its task.

# 4. DESIGN

Groot adopts a modular architecture aimed at fulfilling the requirements outlined in the introduction. Although certain requirements are cross-cutting, our annotations highlight the primarily addressed requirements. Figure 1 depicts Groot's architecture consisting of five central components to tune parameter configurations of applications and their underlying runtime stack. To distinguish between users of the application whose parameters are to be tuned and those utilizing Groot, we refer to the former as users and to the latter as adopters.

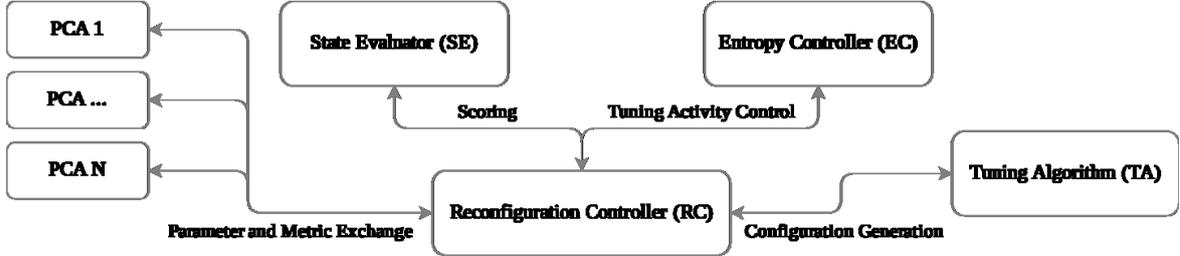

**Figure 1:** Architectural Overview of Groot's components

A Parameter Configuration Agent (PCA) provides a uniform interface to a runtime layer by abstracting parameter setting and metric collection, while also defining tuning objectives and thresholds (e.g., latency) (R2). The Reconfiguration Controller (RC) drives Groot's main loop by orchestrating components: collecting parameters and metrics from PCAs (R1, R4), evaluating and saving the global system state, and issuing new configurations to PCAs. The Entropy Controller (EC) modulates the entropy of proposed configurations over time. The State Evaluator (SE) aggregates PCA information into a global state and integrates layer-specific objectives (R2). The Tuning Algorithm (TA) continuously devises new configurations based on current and past states to improve the overall application performance. (R2, R3). The following sections detail each component's functionality and design decisions.

## Parameter Configuration Agents (PCA)

The Parameter Configuration Agent (PCA) serves as a unified bridge between Groot's central routines and the runtime layers, as shown in Figure 1. Each PCA serves as both sensor and actor: collecting metrics and setting parameter values for a specific runtime layer. As a sensor, it provides reactive, on-demand metrics and parameters with labels specifying tuning objectives, directions, thresholds, and priorities (R2), which the RC uses to guide tuning, e.g., reduce latency with an upper threshold of 200 and priority 3. As an actor, the PCA periodically queries the RC for new configurations and enacts changes, e.g., by updating files or executing queries. We designed the PCA to manage all parameter changes, including handling layer restarts. Thus, Groot supports online tuning (runtime changes) and offline tuning (restarts), as outlined in R3. As with metrics, adopters can label parameters to define ranges and step sizes. Each PCA contributes the respective runtime layer's metric and parameter values to the RC's coherent global state (see Section:RC).

Our modular architecture using PCAs abstracts implementation details of specific runtime layers or technologies therein and, thus, enables Groot to retrieve metrics, discover configuration options, and set new parameters in a technology and application-agnostic way (R5). By further providing a uniform communication interface for PCAs, we enable adopters to implement new PCAs for new layers (R4), e.g., to accommodate for rapidly changing technology. In addition, PCAs may be extended to perform data preprocessing, e.g., sliding-window averaging or aggregation, to adapt to new use cases (R4). Through these design decisions, Groot supports diverse SIV use cases by combining different PCA sets.

## Reconfiguration Controller (RC)

The Reconfiguration Controller (RC) manages Groot's core tasks of observing runtime layers and adjusting parameters by orchestrating PCAs and the Tuning Algorithm (TA) to coordinate data gathering, tuning, and publishing. The RC forwards metrics provided by the PCAs as input to the tuning algorithm and parameter configuration values provided by the tuning algorithm to the PCAs. For that purpose, it queries PCAs for metrics and parameters, discarding partial states to ensure the Tuning Algorithm (TA) always receives a complete system state, which it preprocesses into a compatible format (integer scaling, uniform direction, min/max/step sizes). To stabilize tuning under runtime variability, the RC aggregates several successive states into a snapshot before triggering the TA to propose new parameters, validates them against constraints, and waits a fixed interval to let changes settle. Finally, the RC publishes outputs via uniform interfaces: updated parameter configurations (with values, identifiers, and labels), unified system metrics for monitoring, and runtime statistics for traceability and observability enabling diverse use cases (R1, R4). Furthermore, the RC maintains a history of system states to identify mutation candidates in the tuning algorithm as well as to assess effectiveness of enacted configurations via performance and regression analysis.

To increase comparability between acquired system states and to provide predictable timing during runtime for monitoring and debugging, the RC ensures a stable control loop frequency with a fixed but configurable cycle time. The RC's control loop runs continuously to perform tuning while reacting to changes in workload or system behavior. Thereby, our approach ensures that every optimization step is based on the latest state of all monitored PCAs. In addition, we designed the RC to function as Groot's fixed control entity so that other components may be exchanged or expanded without affecting overall tuning logic.

## State Evaluator (SE)

The State Evaluator (SE) assesses the quality of a configuration by rating the current system state based on observed data and guides the tuning toward higher-performing configurations by comparing the current score to that of previous system states with their configurations. For that purpose, we created the SE to fulfill three main responsibilities:

1. Scoring runtime metrics and aggregating them into system-level performance scores for making parameter configurations comparable (R1).
2. Evaluating performance constraints by weighting multiple objectives (R2).
3. Synthesizing comparable metric values across dynamically observed states.

Our design centralizes performance assessment to capture cross-layer effects of parameter choices, e.g., memory allocation in the containerization layer impacts application runtime performance. To account for such interdependencies, the RC collects parameter values and runtime metrics from all layers via the PCAs.

The SE aggregates metric values collected from PCAs via the RC into a unified system state score, enabling evaluation of combined impacts. Data from PCAs is structured as "metric objects," containing labels (e.g., name=latency) and values (e.g., 300). Labels support identification and classification, distinguishing between tuning metrics (with optimization directions: minimize or maximize) and auxiliary metrics (for profiling or diagnosis).

To support constrained optimization (R2), we introduced three additional labels for tuning goals:

1. A lower threshold for satisfactory performance (e.g., minimum throughput).
2. An upper threshold not to be exceeded (e.g., maximum latency).
3. A weight for relative importance trade-offs (e.g., throughput improvement vs resource reduction).

The SE scores complete system states via weighted sums of tuning metric scores while penalizing violation of thresholds. As viable ranges of parameter values may not be known in advance and exhaustively searching extrema is impractical, we designed Groot to operate with broad parameter ranges: The SE's normalization mechanism continuously observes system state history (maintained by the RC), updates metric extrema and incrementally refines scores. To balance sensitivity and stability, extrema are rounded to scaled halves of the nearest power of ten, avoiding recomputation from minor fluctuations. This ensures normalized scores reflect real workload conditions, avoid distortions from unrealistic assumptions, and remain interpretable. As exploration continues, recalculation frequency decreases as observed boundaries stabilize. On demand recalculations guarantee all states are scored with consistent boundaries, preserving comparability across evaluations. The SE's scoring enables data-driven, adaptive reconfiguration under complex goals without fixed metric boundaries, while avoiding the limitations and preparation overhead of model-based evaluation.

Hence, Groot supports SIVs under multiple constraints or objectives, such as cost efficiency, SLA compliance, or combined goals (R2), e.g., reducing cost with a minimum performance goal may involve minimizing memory and CPU usage while achieving throughput $\geq 5000$ events/s and latency $\leq 500$ ms.

## Tuning Algorithm (TA)

Groot's Tuning Algorithm (TA), invoked by the Reconfiguration Controller (RC), derives new parameter configurations from current and past runtime states. It is an entropy-driven genetic algorithm tailored for resource-intensive, sequential evaluations, where each configuration incurs high cost as values stabilize before the next iteration. Inspired by simulated annealing, the TA evolves one candidate at a time, maintaining a persistent history while supporting high-dimensional tuning (R1). Unlike classical GAs with tournament selection and fixed hyperparameters, it emphasizes early exploration through strong mutations and crossover perturbations, adapts its own hyperparameters, and operates at the gene (single-parameter) level to exploit structural relationships for robust tuning across diverse, high-cost search spaces. Compared to SMBO and Bandit methods, the TA avoids scaling, initialization, and structural issues (Chakraborty & Chakraborty, 2024; Kandasamy et al., 2019; Xu et al., 2025).

The TA integrates Groot's State Evaluator (SE) and Entropy Controller (EC) to regulate adaptation intensity. Entropy balances exploration and exploitation by shaping mutation strength, crossover, re-evaluation, and randomness in selection. Runtime history guides ancestor selection, allowing the TA to combine evaluation results, entropy-weighted variation, and past context to generate new configurations. Unlike typical GAs with synchronous populations and manual tuning of mutation or crossover rates, the TA dynamically adapts these parameters through entropy, eliminating hyperparameter overhead and supporting convergence in high dimensional, real-world settings with variable costs.

Each iteration follows a GA-inspired workflow adapted for entropy control: (1) Ancestor selection ranks candidates by normalized score; (2) A Bernoulli trial, weighted by entropy, decides whether to re-evaluate a past state (exploitation), execute a super-merge of top performers, or proceed with genetic recombination (exploration); (3) Crossover samples parameter values (genes) from two parents during exploration but is disabled during exploitation; (4) Mutation applies either large random changes or small deltas, with number and type governed by entropy; (5) Candidate selection favors random offspring under high entropy and high potential individuals under low entropy.

The TA operates in two entropy-controlled phases. In exploration, crossovers and mutations occur with decaying intensity. At a dynamically positioned inflection point (aligned with search space complexity), the TA transitions to exploitation, reducing crossovers while reusing high-performing states to stabilize around the best configurations and fine-tune promising candidates. In summary, Groot's TA unifies evolutionary dynamics with adaptive entropy control, offering a runtime aware optimization strategy designed for costly evaluations where adaptability and efficiency are critical.

## Entropy Controller (EC)

The Entropy Controller (EC) regulates the randomness of parameter configurations over time by addressing two central challenges in automated optimization: (A) rigid entropy schedules fail to adapt to changing demands, and (B) unpredictable optimization surfaces range from smooth to deceptive or jagged.

Our solution applies three strategies: (1) Varying exploration, higher randomness in early stages and greater stability later, similar to simulated annealing or CMA-ES; (2) varying decay, entropy reduction that adapts to problem complexity, where simpler landscapes converge faster and more complex ones decay more slowly; and (3) externalization, keeping entropy control independent of the tuning algorithm to remain flexible and domain-agnostic. Prior work (Aleti & Moser, 2013) underscores the benefits of adaptive control in evolutionary algorithms for improving their performance.

The EC implements a softened, multi-phase decay function that transitions smoothly from exploration to exploitation. It computes a control variable $\alpha$, proportional to runtime and history size, normalized by the logarithm of search volume and parameter dimensionality. This produces a staircase-like decay curve, where phase changes are dynamically positioned based on telemetry from the Reconfiguration Controller (RC), such as runtime, history size, and search space characteristics. The result is a bounded, context-sensitive entropy schedule that aligns with Groot's genetic algorithm design while remaining compatible with other optimizers.

By modulating entropy through lightweight telemetry rather than manual hyperparameters, the EC enables Groot to explore widely when uncertainty is high and converge reliably when stability is needed. Without requiring prior knowledge of the optimization problem (R5), the EC integrates seamlessly across domains and adjusts its behavior based on inferred complexity and runtime characteristics (R1). The adaptive mechanism provides flexibility, robustness, and efficiency across diverse tuning scenarios without algorithm-specific modifications.

# 5. EVALUATION

To demonstrate that Groot satisfies all posed requirements, we perform an experimental evaluation based on three scenarios rooted in real-world use cases and benchmarks. Each scenario in isolation investigates a subset of the requirements pertaining to characteristics of a specific setup (R1–R3) and all scenarios in conjunction investigate requirements pertaining to different/changing setups (R4–R5) as detailed in Table 1.

**Table 1:** Relation between evaluation scenarios and requirements

|  | R1: Multiple Layers | R2: Multiple Objectives | R3: Offline/Online | R4: Adaptability | R5: Agnostic |
|---|---|---|---|---|---|
| Database | Yes | Yes | Online | Scenarios differ in their layers and utilize different PCAs | Docker Stack |
| Secure Webserver | Yes | Yes | Offline |  | HW TEE |
| Microbenchmark | No | No | Online |  | Simulation |

Our testbed for the evaluation consists of multiple machines: The two hosts for the test application and the load generator consist of Intel Xeon E-2278G CPUs with 16 cores, 32 GB main memory, 40 GbE, M.2 SSD, and run Ubuntu 24.04 (kernel v6.8). Groot is executed on a dedicated server with an Intel Xeon E5-2630 v4 CPU, 16 GB main memory, 10 GbE, a SATA SSD, and runs Ubuntu 24.04 (kernel v6.8). We further use a desktop to facilitate deployments and data collection: a Fujitsu ESPRIMO P957/E90+ desktop machine with an Intel® Core i7–6700 CPU, with 16 GB of RAM, running Ubuntu 22.04 (kernel v5.15).

## Database

The database use case demonstrates Groot's online multi-objective tuning: maximizing performance (throughput, latency) while minimizing resource usage by reducing CPU cores and memory of the database's Docker container (R1-R3). We use the public Postgres v14 (Debian) container with Python added to run a Parameter Configuration Agent (PCA). On a dedicated server, PGbench runs in Docker, initializing Postgres and executing a continuous TPC-B workload (-s 120 -j 95 -c 1005 -P 5 -T 1200 -b tpcb-like).

The database server also hosts a Docker PCA (D-PCA) that adjusts CPU and memory allocations and publishes these metrics to Groot. A kernel PCA (K-PCA) inside the database container exports preselected kernel configuration files (/var/…) as tunable parameters. On the load generator server, a Postgres PCA (PG-PCA) parses PGbench output to provide performance metrics and exposes Postgres parameters and runtime statistics. Groot runs on a third server in an Ubuntu container, collecting metrics and distributing configurations across all three PCAs. In this setup, Groot ingests 20 metrics and outputs 33 parameters, including 22 specific to Postgres.

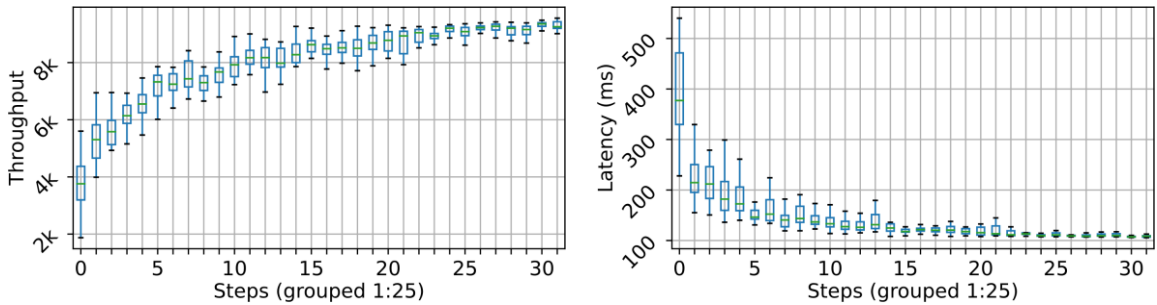

**Figure 2:** Throughput and latency measurements of Postgres during tuning as reported by PGbench.

For the experiment, Groot creates a random configuration for all parameters, then tunes their values over 775 steps per experiment. In Figure 2, we represent each step as the mean across 10 runs. For the visualization, we group 25 steps into one box with the center line indicating the median across the 25 steps. Our results show that the random start state yields a performance with a median throughput of 3707.51 transactions per second and a median latency of 377.15 milliseconds. Groot improves these values to 9274.42 tps and 108.63 ms (median) at the end of the experiments.

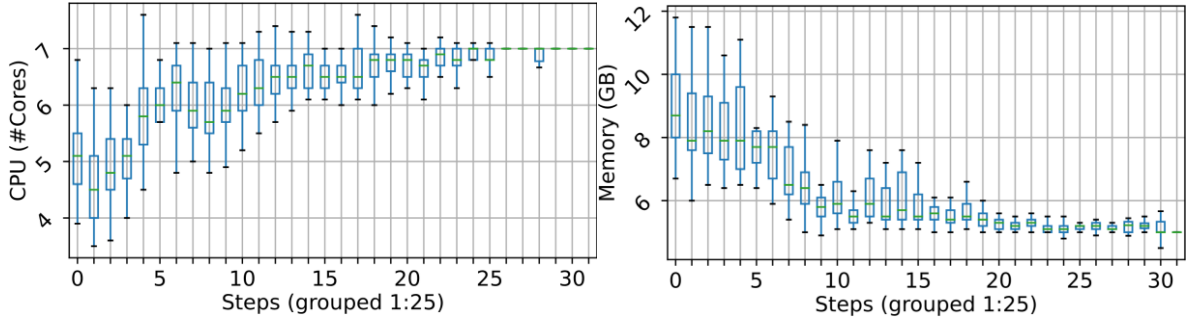
**Figure 3:** Allocation of CPU cores and provided memory to the container running Postgres.

As shown in Figure 3, starting from a median of 5.1 CPU cores, Groot increases CPU allocation during tuning, prioritizing performance over strict resource minimization, while boosting PGbench throughput. The database container was limited to 14 cores and 24 GB of memory, representing a medium-sized, self-hosted setup typical of SIVs. Optimal performance is reached at seven cores, while median memory usage drops from 9.3 GB to 5 GB, the minimum allowed. Our experiment reflects common tuning challenges for small-scale organizations, stressing CPU and memory without large infrastructure. The results demonstrate that Groot autonomously tunes parameters across the runtime stack, improving performance and reducing resource usage without manual guidance, showing that SIVs can achieve significant performance gains and resource savings by applying Groot.

## Secure Webserver

The secure webserver illustrates a Trusted Execution Environment (TEE) scenario, running Nginx inside an Intel-SGX enclave via the SCONE framework (Arnautov et al., 2016) and shows the performance overhead reduction of applications executed in an Intel SGX framework. SCONE exposes TEE specific tunables via environment variables. As changes require restarts, this case highlights Groot's offline tuning. We use SCONE v4.8 and Nginx in an Alpine Docker image. Socat exposes CPU utilization (/proc/stat), alongside Nginx, which is configured for TLS-secured communication. On a second server, wrk2 (commit c4250acb6921) runs in an Alpine container against Nginx. Each run lasts 20 s with 4 threads, 64 connections, and a target rate of 1000 r/s, simulating low demand. We evaluate how Groot reduces resource consumption while tuning latency below a threshold of 100 ms, using a 64 B payload. Alongside Nginx, a Restart-PCA (R-PCA) container adapts SCONE-specific environment variables (SSPINS, SSLEEP, HEAP, ESPINS, SLOTS) and container resources by restarting Nginx with updated settings. Next to wrk2, a W-PCA container reports throughput and latency metrics to Groot but has no tunable parameters. Groot runs on a third server, processing PCA data and providing updated environment variables and resource configurations.

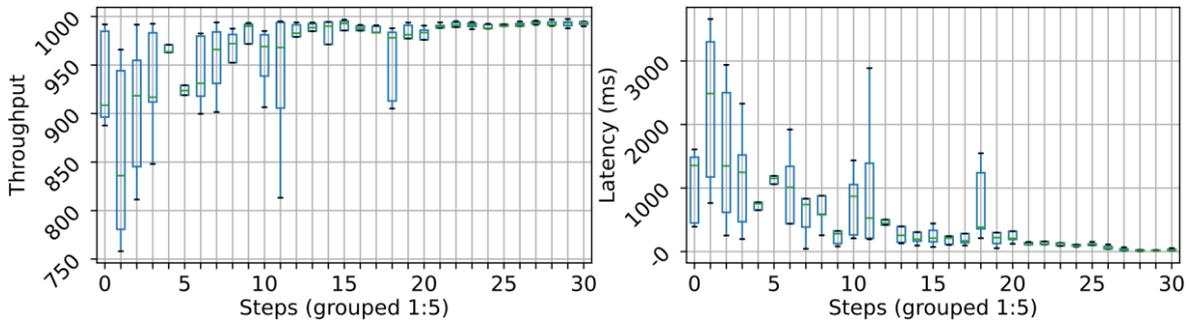
**Figure 4:** Throughput and latency measurements for Nginx measured by wrk2.

Similar to the database experiment, Groot generates random initialization states for each run. In Figure 4, we represent each step as the mean across the 10 runs. For the visualization, we group 5 steps into one box with the center line indicating the median across the 5 steps. Our results show that the random start state yields a performance with a median throughput of 908.6 requests per second and a median latency of 1354.7 milliseconds. Groot improves these values to 994.03 r/s and 18.8 ms at the end of the experiments.

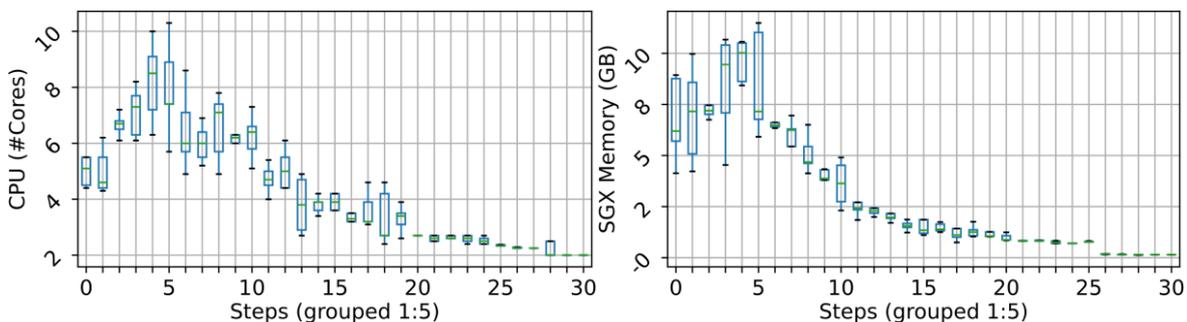

**Figure 5:** Allocation of CPU cores and memory for the Intel SGX enclave executing NGINX.

As shown in Figure 5, starting from a median of 5.1 cores, Groot reduces CPU allocation to 2 cores. Median memory allocation drops from 9.1 GB at experiment start to 2.7 GB. The results of our secure webserver experiment illustrate Groot's effectiveness in reducing resource consumption for applications executed in an Intel SGX framework while tuning the performance.

## Microbenchmark

In a microbenchmark, we evaluate Groot's tuning algorithm (TA) on complex optimization problems, measuring its ability to find near-optimal parameters under varied conditions and how tuning duration scales with scenario complexity (R2, R4). This tests Groot's robustness and general-purpose applicability across large, diverse problem spaces. The search space systematically varies by three factors: number of tunable parameters, number of performance metrics, and parameter value ranges. Per run, the TA optimizes randomly assigned mathematical functions (sum, log, square, product, difference, or average of parameters). Functions are randomly mapped to parameters, creating interdependencies and conflicting objectives. If an experiment requires more than six metrics, functions are randomly reused with new parameter-to-function assignments, generating additional metrics while preserving the system's complexity.

For each configuration, the TA targets 95% of the theoretical maximum performance. The primary outcome is the number of tuning steps required. Each experiment is repeated 1000 times to account for stochastic effects from random initialization, mutation, and crossover. To ensure overall computational feasibility we cap runs at 5000 steps, with fewer than 1% exceeding this limit.

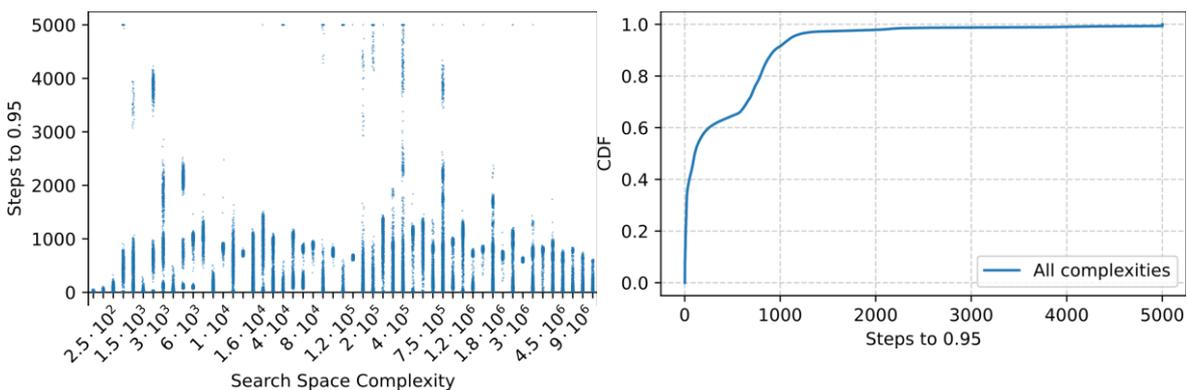

**Figure 6:** Number of steps taken to reach 95% of optimal performance and cumulative distribution.

For the evaluation, the search space complexity is defined as the product of the number of parameters, values per parameter, and metrics. We test the Cartesian product of: parameters [5, 10, 20, 30, 40], metrics [5, 10, 20, 30, 40], and values per parameter [10, 100, 2000, 5000, 10000], using a randomly selected set of mathematical functions. Figure 6 shows that the number of steps to reach 95% of optimal performance remains constant across search space complexities. Furthermore, Figure 6 shows that the TA reaches 95% of optimal performance within 1000 steps for 91.5% of tested complexities. The results indicate that Groot's performance stays consistent across diverse scenarios, including high-complexity search spaces, without prior knowledge, highlighting its adaptability for varied optimization tasks and use cases (R2, R4).

# 6. CONCLUSION

This work presents Groot, a general-purpose configuration tuner capable of tuning on multiple layers for multiple objectives and enacting parameters either online or offline while being domain-agnostic and requiring little knowledge on the tuned system yet discovering effective configurations reliably. Our evaluation shows that Groot meets its design goals in both controlled and realistic scenarios, consistently handling large search spaces, optimizing performance and resources of databases and secure web servers, while supporting online and offline tuning. Groot's properties make it especially well-suited for specialized and innovative ventures (SIVs), who operate under strict cost-performance constraints, manage diverse and evolving software stacks and require parameter tuning to be broadly applicable. Future work will explore automatic detection and classification of changing workload patterns and approaches to decompose the structure of complex search spaces, enabling greater efficiency across domains.